\documentclass[aps,preprint,floats]{revtex4}
\usepackage{graphicx}
\def\lsim{\mathrel {\vcenter {\baselineskip 0pt \kern 0pt
    \hbox{$<$} \kern 0pt \hbox{$\sim$} }}}
\def\gsim{\mathrel {\vcenter {\baselineskip 0pt \kern 0pt
    \hbox{$>$} \kern 0pt \hbox{$\sim$} }}}

\begin{document}

\preprint{hep-ph/0605202}
%\preprint{***}

\title{$B_s-\bar B_s$ Mixing constraints on FCNC and a non-universal $Z^\prime$}

\author{Xiao-Gang He}
\email{hexg@phys.ntu.edu.tw} \affiliation{ Department of Physics,
National Taiwan University, Taipei, Taiwan}

\author{G. Valencia}
\email[]{valencia@iastate.edu} \affiliation{Department of Physics,
Iowa State University, Ames, IA 50011}

\date{\today}

\begin{abstract}

$B_s - \bar B_s$ mixing has been measured recently by D0 and CDF.
The range predicted in the standard model is consistent with data.
However, the standard model central values for $\Delta M_{B_s}$
and also $\Delta M_ {B_d}$ are away from the data which 
may be indications of
new physics. Using the observed values of $\Delta M_{B_s}$ and $\Delta M_{B_d}$ 
we study
general constraints on flavor changing  $Z^\prime$ interactions.
In models with non-universal $Z^\prime$ couplings we find that
significant enhancements over the standard model are still
possible in the rare decay modes $B\to X_s \tau^+ \tau^-$,  $B\to
X_s \nu \bar \nu$ and $B_s\to \tau^+  \tau^-$. Tree level
$Z^\prime$ contributions to $K\to \pi \nu\bar \nu$ are now
constrained to be very small, but one loop effects can still
enhance the standard model rate by a factor of two.

\end{abstract}

\pacs{PACS numbers: 12.15.Ji, 12.15.Mm, 12.60.Cn, 13.20.Eb,
13.20.He, 14.70.Pw}

\maketitle

\section{Introduction}

The D0 and CDF collaborations have recently measured $B_s -\bar
B_s$ mixing with the result \cite{Abazov:2006dm,cdf}
\begin{eqnarray}
&&D0:\;\;\;\;17 ~~ps^{-1} < \Delta M^{exp}_{B_s} < 21 ~~ps^{-1},\nonumber\\
&&CDF:\;\; \Delta M^{exp}_{B_s} = (17.33^{+0.42}_{-0.21}\pm 0.07)
~~ps^{-1}.
\end{eqnarray}
This last measurement is sufficiently precise to place new
constraints on tree-level flavor changing neutral currents. In
this paper we explore the consequences of these constraints in
models with an additional gauge boson, a $Z^\prime$. We first
discuss general constraints and then specialize to the case of
non-universal $Z^\prime$ models. For our numbers we will use the
CKMfitter \cite{Charles:2004jd} average including these
measurements,
\begin{equation}
\Delta M^{exp}_{B_s} = (17.34^{+0.49}_{-0.20}) ~~ps^{-1}.
\end{equation}

This mass difference is related to the $B_s - \bar B_s$ mixing
parameter $M^{B_s}_{12}$ by $\Delta M_{B_s} = 2 |M^{B_s}_{12}|$,
if the lifetime difference is neglected. In the Standard Model,
$M^{B_s}_{12}$ arises from the so called ``box'' diagram and is
given by
\begin{eqnarray}
&&M^{B_s,SM}_{12} = {G_F^2\over 12 \pi^2} \eta_B m_{B_s}
\xi^2 f_{B_d}^2 B_{B_d} m^2_W S(x_t) (V_{ts}V^*_{tb})^2,\nonumber\\
&&S(x) = {4x-11x^2+x^3\over 4(1-x)^2} -{3 x^3\ln x\over 2
(1-x)^3},\label{SM}
\end{eqnarray}
where $x_t = m^2_t/m_W^2$, $\eta_B = 0.551\pm 0.007$ is a QCD
correction. The hadronic parameters $f_{B_d} = (0.191\pm 0.027)$
GeV, $B_{B_d} =1.37\pm 0.14$ and $\xi =
f_{B_s}\sqrt{B_s}/f_{B_d}\sqrt{B_d} =1.24\pm 0.04 \pm 0.06$ are
obtained from lattice calculations \cite{Charles:2004jd}.  This
value of $f_{B_d}$ is in reasonable agreement with the recently
observed branching ratio  $B(B_d \to \tau \nu_\tau) =
(1.06^{+0.3}_{-0.28}(sat)^{+0.18}_{-0.16}(syst))\times 10^{-4}$
\cite{Ikado:2006un}.

To quantify the uncertainty in the input parameters to
Eq.~(\ref{SM}), we use the latest result from the CKMfitter overall
fit (excluding the measurement) \cite{Charles:2004jd},
\begin{equation}
\left(\Delta M_{B_s}\right)_{SM} = 21.7^{+13.1}_{-9.1} ~~ps^{-1}
\end{equation}
where the errors indicate the ${\mathbf 3\sigma}$ range.  Notice
that the central value of this prediction is slightly higher than
the measured mass difference, although the predicted and measured
ranges are in good agreement.

This agreement between the SM prediction and the data places
stringent constraints on new physics that will become more severe
as the theoretical uncertainty is reduced. There are many models
beyond the SM containing additional flavor changing sources  which
can be constrained by recent data on $\Delta
M_{B_s}$~\cite{Group1}. We concentrate here on the impact
of the measured $\Delta M_{B_s}$ on non-universal $Z'$ models that are
motivated by the apparent anomaly in the measurement of $A^b_{FB}$
at LEP~\cite{He:2002ha,He:2003qv,He:2004it}. These models are
variations of left-right models in which the right-handed
interactions single out the third generation with enhanced
couplings to the $Z^\prime$.

\section{Generic bounds on new physics}

\begin{figure}[htb]
\begin{center}
\includegraphics[width=8cm]{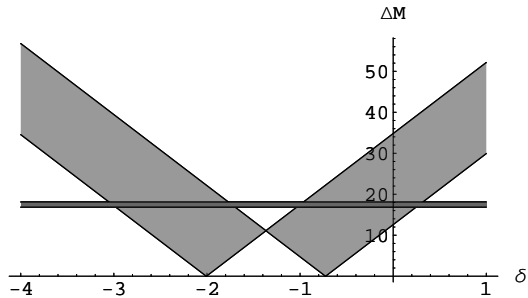}
\end{center}
\caption{$\Delta M_{B_s}$ as a function of $\delta_{B_s}$ for a
new physics contribution assumed to be real. We show a shaded band
obtained by adding the new physics to the 3-$\sigma$ standard
model range. The horizontal band corresponds to the 3-$\sigma$
experimental range. }\label{range}
\end{figure}

We begin by considering a generic new physics  contribution to
$M^{B_s,N}_{12}$ that is real, and add it to the standard model
3-$\sigma$ range. By requiring this to overlap with the 3-$\sigma$
range for the measured   $\Delta M_{B_s}$ we extract the allowed
range for new physics. Normalized to the central value of the
 measured   $\Delta M_{B_s}$ we find
\begin{eqnarray}
\delta_{B_s} \equiv {2 M^{B_s,N}_{12}\over \Delta M^{exp}_{B_s}}
\sim (-3 {\rm ~to~} -1.7)~{\rm ~or~}(-1 {\rm ~to~} 0.27),
\end{eqnarray}
as seen in Figure~(\ref{range}).

Because the central value of the SM prediction is already larger
than the measured mass difference, there is little room for a new
contribution in phase with the standard model. A larger range is
allowed for a new physics contribution opposite in sign to the SM.
More generally, $\delta_{B_s}$ is complex and we refer to  cases
in which  the real part of $\delta_{B_s}$ has the same sign as
(opposite sign to) the SM as having constructive (destructive)
interference with the  SM.

\begin{figure}[htb]
\begin{center}
\includegraphics[width=5cm]{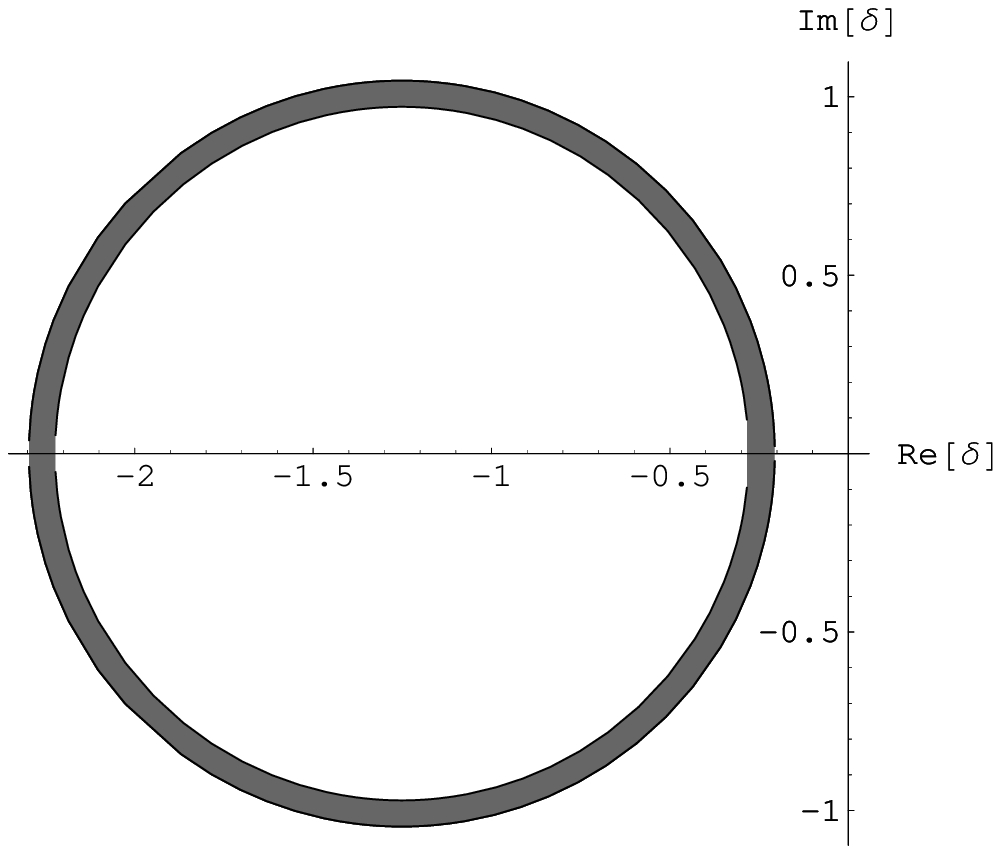}
\includegraphics[width=8cm]{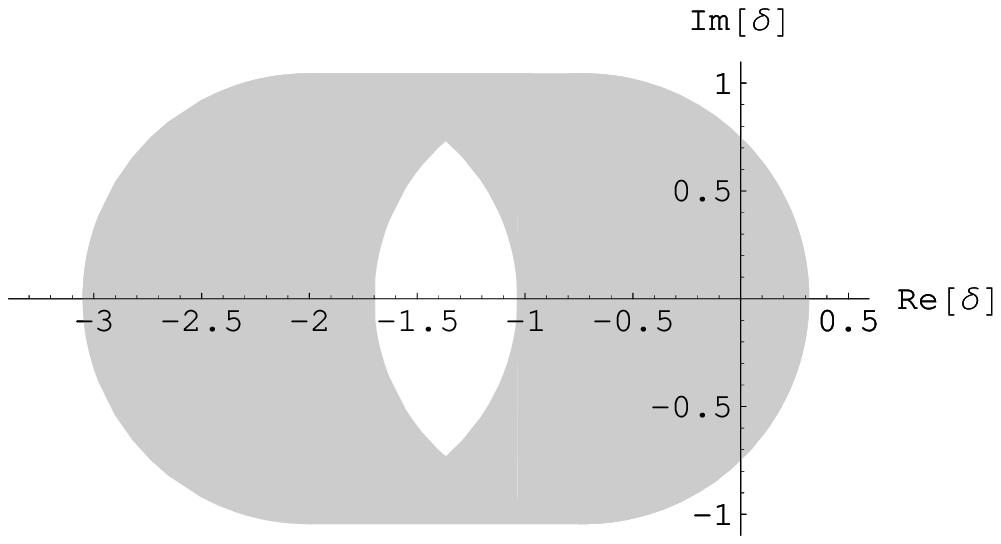}
\end{center}
\caption{Constraints on $Re(\delta_{B_s})$ and $Im(\delta_{B_s})$
for a complex new physics contribution to $M_{12}^{B_s}$. The
shaded regions indicate where the prediction falls within the
3-$\sigma$ experimental range on $\Delta M_{B_s}$ when we use (a)
the central value of the SM prediction; and (b) the 3-$\sigma$
range for the SM prediction. }\label{bound}
\end{figure}

For a new physics contribution that is complex, we require that
$2|M^{B_s,N}_{12} + M^{B_s,SM}_{12}|$ reproduce the measured mass
difference. Once again we allow a 3-$\sigma$ range in both the SM
prediction and the measurement. In Figure~(\ref{bound}) we show
the allowed ranges for $Re(\delta_{B_s})$ and $Im(\delta_{B_s})$
for two cases. In the first case we use the central value of the
SM prediction, $\Delta M_{B_s} = 21.7 ~~ps^{-1}$ whereas in the
second case we use the full 3-$\sigma$ range $\Delta M_{B_s} =
12.6 ~~ps^{-1}{\rm ~~to~~} 34.8~~ps^{-1}$. We see again that there
is very little room for a constructive new physics contribution.

\section{Constraints on generic $Z^\prime$ models}

We now restrict ourselves to the case where the new physics
contributions originate in the exchange of $Z^\prime$ bosons with
flavor changing couplings. The flavor changing $Z^\prime$
couplings can be written  in general as
\begin{eqnarray}
{\cal L} = {g\over 2 c_W} \bar q_i \gamma^\mu (a_{ij}P_L +
b_{ij}P_R )q_j Z^\prime_\mu\;.
\end{eqnarray}

A tree-level exchange of the $Z^\prime$ generates the effective
Lagrangian responsible for neutral meson mixing (and in particular
 $B_s -\bar B_s$ mixing) at the $M_{Z^\prime}$ scale,
\begin{eqnarray}
{\cal L}_{Z^\prime} = - {G_F\over \sqrt{2}} {M^2_Z\over
M^2_{Z^\prime}} [a^2_{ij} O_{LL} + b^2_{ij}O_{RR} +
2a_{ij}b_{ij}O_{LR}],
\end{eqnarray}
where the  operators are given by
\begin{eqnarray}
&&O_{LL} =\bar q_i \gamma^\mu P_L q_j \bar q_i\gamma_\mu P_L q_j,
\;\;O_{RR} =\bar q_i \gamma^\mu P_R q_j \bar q_i\gamma_\mu P_R
q_j,\nonumber\\
&&O_{LR} =\bar q_i \gamma^\mu P_L q_j \bar q_i\gamma_\mu P_R q_j,
\;\;\tilde O_{LR} =\bar q_i P_L q_j \bar q_iP_R q_j.
\end{eqnarray}

The operator $\tilde O_{LR}$ does not appear directly in
$Z^\prime$ exchange, but is induced by renormalization through
mixing with $O_{LR}$.  Starting from an effective Lagrangian at
the high energy scale $m$ given by
\begin{eqnarray}
{\cal L} = a_{LL}(m)O_{LL}+a_{RR}(m)O_{RR} + a_{LR}(m)O_{LR} +
\tilde a_{LR}(m) \tilde O_{LR},
\end{eqnarray}
we find at a low energy scale $\mu = m_b$ relevant to $B_s$
mixing,
\begin{eqnarray}
{\cal L} = a_{LL}(\mu)O_{LL}+a_{RR}(\mu)O_{RR} + a_{LR}(\mu)O_{LR}
+ \tilde a_{LR}(\mu) \tilde O_{LR}.
\end{eqnarray}
At leading order in QCD RG running, the coefficients are
\cite{Ecker:1985ei,Barger:2003hg}
\begin{eqnarray}
&&a_{LL}(\mu) = a_{LL}(m)\eta_{LL}(\mu),\;\;a_{RR}(\mu) =
a_{RR}(m)\eta_{RR}(\mu),\nonumber\\
&&a_{LR}(\mu) = a_{LR}(m)\eta_{LR}(\mu),\;\;\tilde a_{LR}(\mu) =
\tilde
a_{LR}(m)\tilde \eta_{LR}(\mu) + {2\over 3} a_{LR}(m)(\eta_{LR}-\tilde \eta_{LR}),\nonumber\\
 &&\eta_{LL}(\mu) =
\left (\eta_m\right )^{6/23},\;\;\eta_{RR}(\mu) = \left (\eta_m
\right )^{6/23},\nonumber\\
&&\eta_{LR}(\mu) = \left ( \eta_m \right )^{3/23},\;\;\tilde
\eta_{LR}(\mu) = \left (\eta_m \right )^{-24/23}.
\end{eqnarray}
where $\eta_m \equiv \alpha_s(m)/\alpha_s(\mu)$.

From the low energy effective Lagrangian one obtains the mass
difference in terms of the ``bag factors'',
\begin{eqnarray}
&&M^{P,Z'}_{12} = -{1\over 3} f_P^2 m_P B_p\left [ a_{LL}(\mu) +
a_{RR}(\mu)+ a_{LR}(\mu) (-{3\over 4} + {\epsilon\over 2}) +
\tilde a_{LR}(\mu) ({1\over 8} - {3\epsilon\over 4} )\right ],
\end{eqnarray}
where $B_P=B_{LL}=B_{RR}=B_{LR}$ is the ratio between the matrix
element  $<P|\bar q\gamma^\mu \gamma_5 b \bar q \gamma_\mu
\gamma_5 b|P>$ and its value in factorization. Similarly,
$\epsilon$ is defined as $\epsilon = (\tilde
B_{LR}/B_{LL})(m^2_P/(m_i+m_j)^2)$ where $\tilde B_{LR}$ is the
ratio between the matrix element $<P|\bar q \gamma_5 b \bar q
\gamma_5 b|P>$ and its value in factorization. We will use
$\epsilon =1$ for our numerical results.

With all this we finally obtain the new physics contribution to
$M_{12}$ from $Z^\prime$ exchange,
\begin{eqnarray}
M^{P,Z^\prime}_{12} &=& {G_F\over \sqrt{2}}{m^2_Z\over
m^2_{Z^\prime}}\eta_{Z^\prime}^{6/23}{1\over 3} f^2_P M_P B_P
\left ( a^2_{ij}
+ b^2_{ij}\right. \nonumber\\
& +&\left . \eta_{Z^\prime}^{-3/23} {1\over 2}
a_{ij}b_{ij}(2\epsilon -3) + {2\over
3}(\eta_{Z^\prime}^{-3/23}-\eta_{Z^\prime}^{-30/23}){1\over
4}a_{ij}b_{ij} (1-6\epsilon)\right ).
\end{eqnarray}

The mass difference $\Delta M^P$ is then obtained by adding the SM
and new physics contributions,
\begin{eqnarray}
\Delta M^P = 2 \left|M^{P,SM}_{12}+M^{P,Z^\prime}_{12}\right|.
\end{eqnarray}

\begin{figure}[htb]
\begin{center}
\includegraphics[width=6cm]{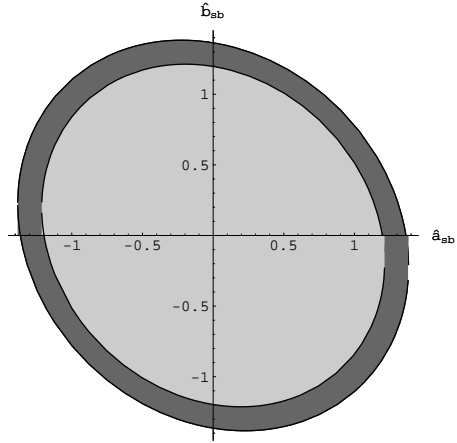}
\end{center}
\caption{Constraints on the flavor changing parameters
$\hat{a}_{sb}\equiv (m_Z/m_{Z^\prime})a_{sb}\times 10^3$ and
$\hat{b}_{sb}\equiv (m_Z/m_{Z^\prime})b_{sb}\times 10^3$. The
shaded region is allowed when the SM contribution varies in its
3-$\sigma$ range. The darker shaded region indicates corresponds
to $(\Delta M_{B_s})_{SM}=12.6~~ps^{-1}$. }\label{constraint}
\end{figure}

As mentioned before, the new physics contribution can be
constructive or destructive with respect to the SM. In the case of
$B_s$ mixing, if $a_{sb}$ and $b_{sb}$  are real relative to the
CKM matrix element $V_{tb}^*V_{ts}$, the new contributions are
constructive. We show in Figure~(\ref{constraint}) the allowed
region for the parameters $a_{sb}$, and $b_{sb}$ (assumed to be
real). The shaded region is obtained
for a SM contribution allowed to vary in its 3-$\sigma$ range. For
emphasis, the darker shaded region corresponds to $(\Delta
M^{B_s})_{SM}= 12.6 ~~ps^{-1}$ indicating the lower end of the
theory range. Notice that for the central value of the SM
prediction, $(\Delta M^{B_s})_{SM}= 21.7 ~~ps^{-1}$, there is no
room for constructive new physics.

\begin{figure}[htb]
\begin{center}
\includegraphics[width=6cm]{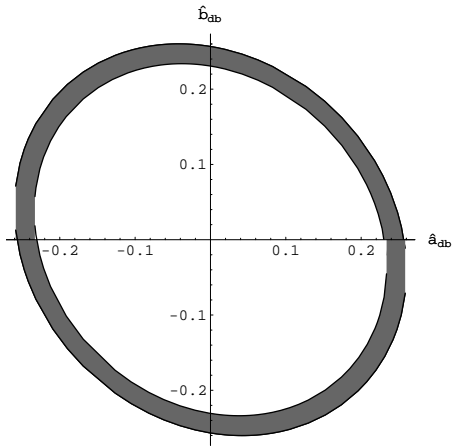}
\end{center}
\caption{Constraints on the flavor changing parameters
$\hat{a}_{db}\equiv (m_Z/m_{Z^\prime})a_{db}\times 10^3$ and
$\hat{b}_{db}\equiv (m_Z/m_{Z^\prime})b_{db}\times 10^3$. The
shaded region indicates the 3-$\sigma$ experimental range on
$\Delta M_{B_d}$ corresponding to the central value of the SM
prediction. }\label{rangebd}
\end{figure}

A similar exercise can be done for $B_d$ mixing with the
3-$\sigma$ SM range from Ref.~\cite{Charles:2004jd} as well as the
HFAG experimental average \cite{hfag}
\begin{eqnarray}
\left(\Delta M_{B_d}\right)_{SM} &=& 0.394^{+0.361}_{-0.162} ~~ps^{-1}\nonumber \\
\left(\Delta M_{B_d}\right)_{exp}&=& (0.507 \pm 0.004) ~~ps^{-1}.
\end{eqnarray}
Assuming that the new physics is real it will also interfere
constructively with the SM. Taking the central value of the SM
prediction and requiring the total $\Delta M^{B_d}$ to fall within
the 3-$\sigma$ experimental range leads to the allowed region
shown in Figure~(\ref{rangebd}).

\section{A non-universal
$Z^\prime$ model and flavor changing parameters}

We now apply the above constraints  to a $Z'$ model with
non-universal couplings. The non-universal $Z^\prime$ models we
consider here have been discussed in
Ref.~\cite{He:2002ha,He:2003qv,He:2004it} motivated by the
apparent anomaly in the measurement of $A^b_{FB}$ at LEP
\cite{Chanowitz:2001bv,Abbaneo:2001ix}. The models are variations
of left-right models in which the right-handed interactions single
out the third generation giving enhanced $Z^\prime$ couplings to
the $b$ and $t$ quarks and to the $\tau$ and $\nu_\tau$ leptons.
In general the models contain tree-level  flavor changing neutral
currents as well. We find that the new measurement of $\Delta
M_{B_s}$ can place stringent constraints on some of the parameters
of the model, but that there is still room for a substantial
enhancement in the modes $B\to X_s \tau^+ \tau^- (\nu_\tau \bar
\nu_\tau )$, $B_s\to \tau^+  \tau^-$, and also $K\to \pi \nu_\tau
\bar \nu_\tau$. We briefly review the relevant aspects of the
models and refer the reader to
Ref.~\cite{He:2002ha,He:2003qv,He:2004it} for details.

In these models the first two generations are chosen to have the
same transformation properties as in the standard model with
$U(1)_Y$ replaced by $U(1)_{B-L}$,
\begin{eqnarray}
&&Q_L = (3,2,1)(1/3),\;\;\;\;U_R = (3,1,1)(4/3),\;\;\;\;D_R =
(3,1,1)(-2/3),
\nonumber\\
&&L_L = (1,2,1)(-1),\;\;\;\;E_R = (1,1,1)(-2). \label{gens12}
\end{eqnarray}
The numbers in the first parenthesis are the $SU(3)$, $SU(2)_L$
and $SU(2)_R$ group representations respectively, and the number
in the second parenthesis is the $U(1)$ charge. For the first two
generations the $U(1)$ charge is the same as the $U(1)_Y$ charge
in the SM and for the third generation it is the usual
$U(1)_{B-L}$ charge of LR models. The third generation is chosen
to transform differently,
\begin{eqnarray}
&&Q_L(3) = (3,2,1)(1/3),\;\;\;\;Q_R(3) = (3,1,2)(1/3),\nonumber\\
&&L_L(3) = (1,2,1)(-1),\;\;\;\;L_R = (1,1,2)(-1). \label{gen3}
\end{eqnarray}

The correct symmetry breaking and mass generation of particles can
be induced by the vacuum expectation values of three Higgs
representations: $H_R = (1,1,2)(-1)$, whose non-zero vacuum
expectation value (vev) $v_R$ breaks the group down to
$SU(3)\times SU(2)\times U(1)$; and the two Higgs multiplets, $H_L
= (1,2,1)(-1)$ and $\phi = (1,2,2)(0)$, which break the symmetry
to $SU(3)\times U(1)_{em}$.

The models contain flavor changing neutral currents at tree level
that contribute to $B_s$ mixing and other related flavor changing
decays,  the relevant interactions are \cite{He:2003qv},
\begin{eqnarray}
{\cal L}_Z &=&  {g\over 2}\tan\theta_W (\tan\theta_R+\cot\theta_R)
(\sin\xi_Z Z_\mu + \cos\xi_Z Z^\prime_\mu) \nonumber \\
&\times &\left( \bar d_{Ri} \gamma^\mu V^{d*}_{Rbi} V^{d}_{Rbj}
d_{Rj} -\bar u_{Ri} \gamma^\mu V^{u*}_{Rti} V^{u}_{Rtj} u_{Rj}
+\bar \tau_R \gamma^\mu \tau_R -\bar \nu_{R \tau} \gamma^\mu
\nu_{R \tau} \right) \label{nmcoups}
\end{eqnarray}
In this expression $g$ is the usual $SU_L(2)$ gauge coupling,
$\theta_W$ the usual electroweak angle, $\theta_R$ parametrizes
the relative strength of the right-handed interactions, $\xi_Z$ is
the $Z$-$Z^\prime$ mixing angle and $V^{u,d}_{Rij}$ are the
unitary matrices that rotate the right-handed up-(down)-type
quarks from the weak eigenstate basis to the mass eigenstate basis
\cite{He:2003qv}.

The relative strength of left- and right-handed interactions is
determined by the parameter $\cot\theta_R$. In the limit in which
this parameter is large, the new right-handed interactions affect
predominantly the third generation. It was found in
Ref.~\cite{He:2003qv} that the measurement of $g_{R\tau}$ at
LEP\cite{Abbaneo:2001ix} implies a small $\cot\theta_R \xi_Z \leq
10^{-3}$ if the new interaction affects the third generation
leptons as well as the quarks. It is possible to construct models
in which the third generation lepton couplings are not enhanced.
Here we consider models in which they are enhanced but in which
the $Z-Z^\prime$ mixing is negligible.

In Ref.~\cite{He:2003qv}, the process $e^+e^- \rightarrow b
\bar{b}$ at LEP-II was used to obtain a lower bound for the mass
of the new $Z^\prime$ gauge boson for a given $\cot\theta_R$. For
our present purpose that bound can be approximated by the relation
\begin{equation}
\cot\theta_R \tan\theta_W \left({M_W \over M_{Z^\prime}}\right)
\sim 1. \label{appbound}
\end{equation}

Within this framework there are two potentially large sources of
FCNC. The first one, through the coupling $\bar{d}_i\gamma_\mu P_R
d_j Z^{\prime \mu}$ which occurs at tree level and which also
receives large one-loop corrections (enhanced by $\cot\theta_R$).
There
is a second operator responsible for FCNC which has the form
$\bar{d}_i\gamma_\mu P_L d_j Z^{\prime \mu}$. This operator first
occurs at one-loop with a finite coefficient that is enhanced by
$\cot\theta_R$, and is present even when there are no
FCNC at tree-level. Because it is enhanced by $\cot\theta_R$, it
can contribute to a low energy FCNC process at the same level as
the ordinary electroweak penguins mediated by the $Z$ boson even
though $M_{Z^\prime} >> M_Z$.  It can be written as
\cite{He:2004it}
\begin{equation}
{\cal L}_{eff} = {g^3 \over 16 \pi^2} \tan\theta_W\cot\theta_R
V^\star_{ti} V_{tj} I(\lambda_t,\lambda_H) \bar{d}_i \gamma_\mu
P_L d_j\  Z^{\prime \mu} \label{effl},
\end{equation}
where $I(\lambda_t,\lambda_H)$ ($\lambda_i = m^2_i/m^2_W$) is the
corresponding Inami-Lim type function. With a Higgs mass in the
range of a few hundred GeV, this function varies between a few and
about 20 \cite{He:2004it}. When a third generation lepton pair is
attached to the $Z^\prime$ in Eq.~(\ref{effl}), a second factor
$\cot\theta_R$ is introduced which compensates for the small
$M_Z/M_{Z^\prime}$ ratio and makes this mechanism comparable to
the standard $Z$ penguin as follows from Eq.~(\ref{appbound}).

Collecting the above FCNC interactions, we find that for large
$\cot\theta_R$, $Z^\prime$ exchange will produce the following
effective flavor changing parameters,
\begin{eqnarray}
&&a_{ij} = {\alpha\over 2 \pi \sin^2\theta_W}I(\lambda_t,
\lambda_H) \cos\theta_W \tan\theta_W \cot\theta_R
V^*_{ti}V_{tj},\nonumber\\ &&b_{ij} = \cos\theta_W \tan\theta_W
\cot\theta_R \cos\xi_Z V^{d*}_{bi}V^d_{bj}.
\end{eqnarray}
Here $\cos\xi_Z = 1$ since we are working in the limit of no
$Z-Z^\prime$ mixing.

It is interesting to find  that the one loop generated $a_{ij}$
can have a significant contribution to $B_s$ mixing. For example,
with $\cot\theta_R \tan\theta_W (m_W/m_{Z^\prime} )= 1$ and
$b_{bs} = 0$, Figure~(\ref{constraint}) shows that the range
$0.0012 \lsim (m_Z/ m_{Z^\prime})|a_{bs}| \sim 0.0015$ added to
the lowest bound of the SM reproduces the measured $B_s$ mass
difference. This range implies
\begin{eqnarray}
5.5 \lsim I(\lambda_t, \lambda_H)\left|
\frac{V_{tb}^*V_{ts}}{0.04}\right| \lsim 6.5 \label{intran}
\end{eqnarray}
This range is accessible in our models with reasonable parameters
for the Higgs mass and $\cot\theta_R$ (See Figure 3 in
Ref.~\cite{He:2004it}). Interestingly, this is the same range in
which $K^+ \to \pi^+ \nu \nu$ reproduces the branching ratio
measured by E787 and E949 (which requires $I(\lambda_t,\lambda_H)
= 5.54$ \cite{He:2004it}).

If we take the value, $I(\lambda_t,\lambda_H) = 5.54$, then $(m_Z/
m_{Z^\prime})|a_{bs}| \sim 0.0012$ and from
Figure~(\ref{constraint}) we see that the following range of
values is allowed for real $b_{bs}$
\begin{eqnarray}
0.0005\sim \left|V^{d*}_{Rbb}V^d_{Rbs}\right| \sim 0.0009.
\end{eqnarray}

\begin{figure}[htb]
\begin{center}
\includegraphics[width=8cm]{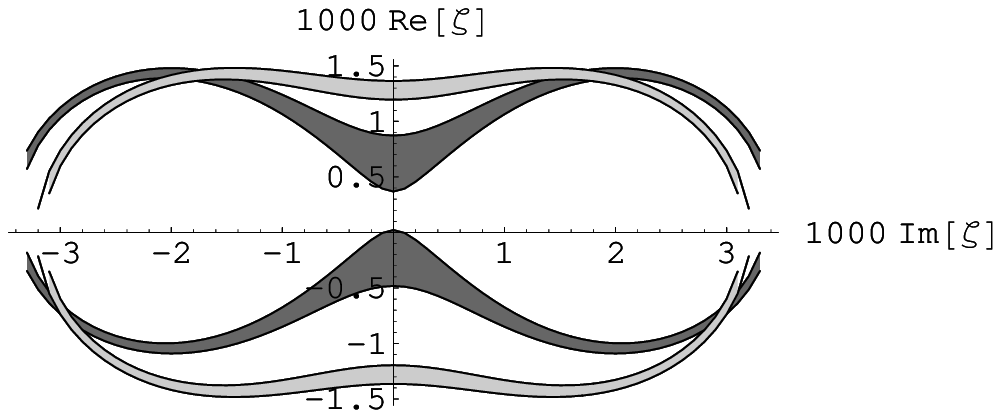}{\hspace{0.5in}}
\includegraphics[width=6cm]{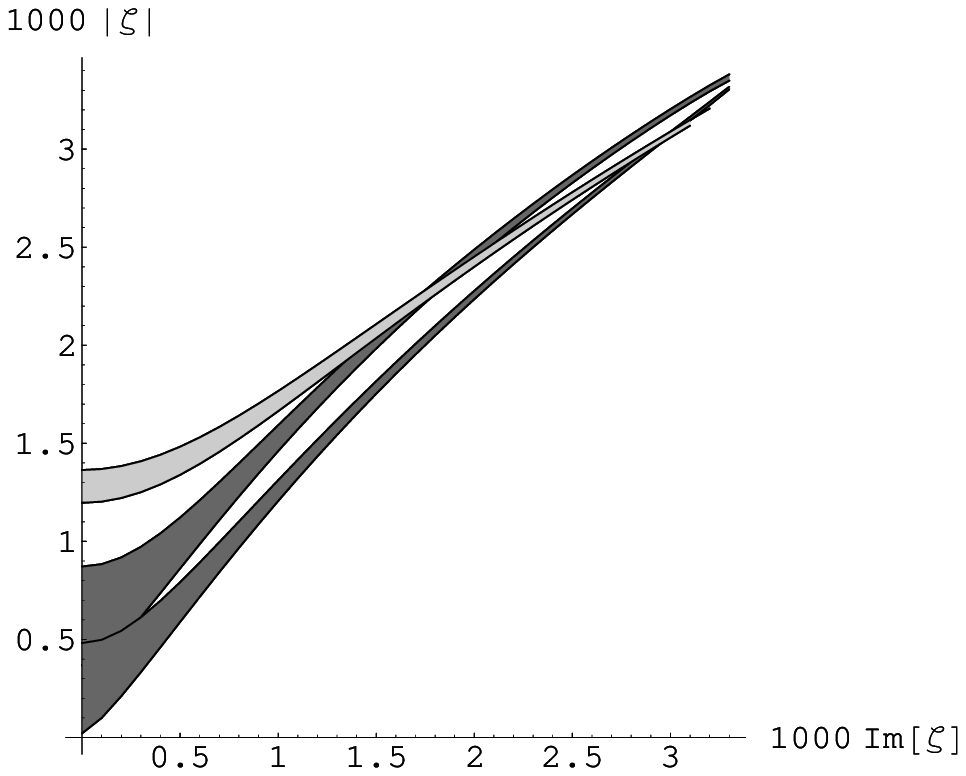}
\end{center}
\caption{ Constraints on the flavor changing parameter
$\zeta\equiv V^{d*}_{bb}V^d_{bs}$ with $\Delta
M^{SM}_{B_s}=12.6ps^{-1}$. The shaded regions correspond to the
new physics contribution necessary to match the 3-$\sigma$ $\Delta
M_{B_s}$ range for $a_{bs}=0$ (light gray) and $a_{bs}=-0.0012$
(dark gray).}\label{parameter}
\end{figure}

\begin{figure}[htb]
\begin{center}
\includegraphics[width=8cm]{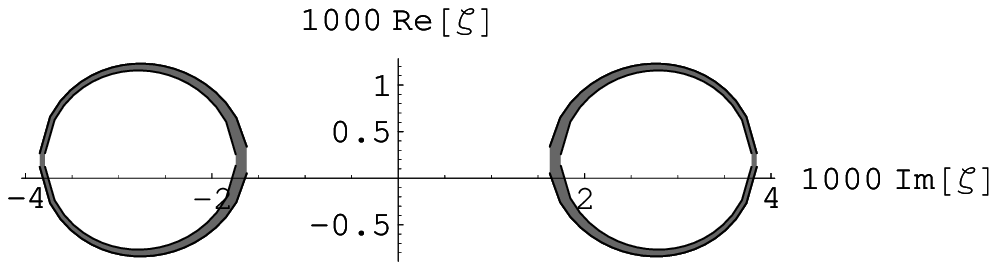}{\hspace{0.5in}}
\includegraphics[width=6cm]{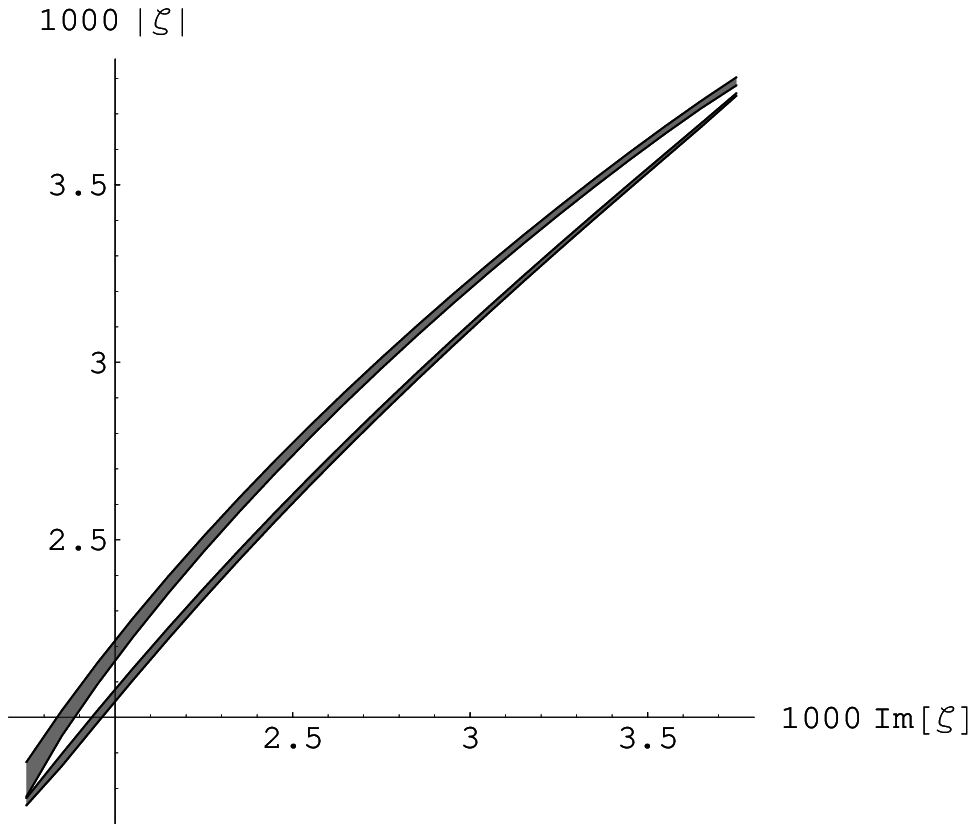}
\end{center}
\caption{ Constraints on the flavor changing parameter
$\zeta\equiv V^{d*}_{bb}V^d_{bs}$ with $\Delta M^{SM}_{B_s} =
21.7ps^{-1}$. The shaded regions correspond to the new physics
contribution necessary to match the 3-$\sigma$ $\Delta M_{B_s}$
range when $a_{bs}=-0.0012$.}\label{parameterp}
\end{figure}

Allowing $V^{d*}_{bb}V^d_{bs}$ to be complex, we find an allowed
range shown in Figures~(\ref{parameter}) and ~(\ref{parameterp}).
In Figure~(\ref{parameter}) we use $\Delta
M^{SM}_{B_s}=12.6ps^{-1}$, at the low end of the theoretical
range. The region shaded in light gray corresponds to $a_{bs}=0$,
whereas the region in dark gray corresponds to $a_{bs}=-0.0012$.
If we take  $a_{bs}=-0.0012$ but use the central value of the SM
range then we obtain Figure~(\ref{parameterp}). Notice that in
this case there are no solutions for real values of $b_{bs}$ (or
$\zeta$). New physics in this case is only allowed with a large CP
violating phase.

\section{$B\to X_s \tau^+  \tau^- (\nu \bar \nu )$,
$B_s \to \tau^+  \tau^-$ and $K\to \pi\nu\bar \nu$ }

We now show that the constraints on the flavor changing parameters
from $\Delta M_{B_s}$ still allow for a substantial enhancement in
$b \to s \tau^+ \tau^- (\nu_\tau \bar \nu_\tau)$ transitions. In
the large $\cot \theta_R$ limit, a $Z^\prime$ exchange at tree
level leads to an effective interaction
\begin{equation}
{\cal L} = {g^2\tan^2\theta_W\cot^2\theta_R\over 4 M^2_{Z^\prime}}
V^{d\star}_{Rbs}V^{d}_{Rbb}\bar s \gamma_\mu P_R b \
(\bar{\nu}_\tau\gamma^\mu P_R \nu_\tau - \bar \tau \gamma^\mu P_R
\tau) + h.c. \label{tleffl},
\end{equation}
and at one loop level to
\begin{eqnarray}
{\cal L} = {g^2\tan^2\theta_W\cot^2\theta_R\over 4 M^2_{Z^\prime}}
{g^2 \over 8 \pi^2} V^*_{ts}V_{tb}I(\lambda_t,\lambda_H) \bar s
\gamma_\mu P_L b \ (\bar{\nu}_\tau\gamma^\mu P_R \nu_\tau - \bar
\tau \gamma^\mu P_R \tau) + h.c.
\end{eqnarray}

The corresponding transitions in the SM are mediated by the
effective Hamiltonian
\begin{equation}
\tilde{H}_{eff} = {G_F \over \sqrt{2}}{2\alpha\over \pi
\sin^2\theta_W} V^\star_{ts}V_{tb} \bar{s}\gamma_\mu P_L b \
\left[X(x_t) \sum_\ell \bar{\nu_\ell}\gamma^\mu P_L \nu_\ell -
Y(x_t) \bar \tau \gamma^\mu P_L \tau\right]+ h.c., \label{comp}
\end{equation}
where $x_t =m_t^2/M_W^2$ and the Inami-Lim functions $X(x_t)$ and
$Y(x_t)$ are  approximately equal to 1.6 and 1.06 respectively
\cite{Buchalla:1995vs}.

Comparing the tree-level $Z^\prime$ exchange and SM contributions,
we have
\begin{eqnarray}
&&{\Gamma_{new}(B\to X_s \nu\bar \nu)\over \Gamma_{SM}(B\to X_s
\nu\bar \nu)} \approx  1130 \cot^4\theta_R \tan^4\theta_W
\left({M_W \over M_{Z^\prime}}\right)^4
\left|{V^{d\star}_{Rbs}V^{d}_{Rbb}\over
V^\star_{ts}V_{tb}}\right|^2,\nonumber\\
&&{\Gamma_{new}(B_s\to \tau \bar \tau)\over \Gamma_{SM}(B_s\to
\tau \bar \tau)} \approx  7730 \cot^4\theta_R \tan^4\theta_W
\left({M_W \over M_{Z^\prime}}\right)^4
\left|{V^{d\star}_{Rbs}V^{d}_{Rbb}\over
V^\star_{ts}V_{tb}}\right|^2. \label{kpnnrat}
\end{eqnarray}
In the SM, with $|V_{ts}^*V_{tb}|\approx 0.04$ these branching
ratios are predicted to be  $B(B\to X_s \nu \bar \nu) = 4\times
10^{-5}$ and $B(B_s \to \tau \bar \tau) = 1.1\times 10^{-6}$.

For $B\to X_s \tau^+  \tau^-$ it is easier to compare the new
contribution to the semileptonic decay,
\begin{equation}
\frac{\Gamma_{new}(B\to X_s \tau \bar \tau)}{\Gamma_{SM}(B \to X_c
e^-\bar{\nu})} \approx 0.06 \cot^4\theta_R \tan^4\theta_W
\left({M_W \over
M_{Z^\prime}}\right)^4\left|{V^{d\star}_{Rbs}V^{d}_{Rbb}\over
V_{cb}}\right|^2
\end{equation}
The short distance contributions to $B\to X_s \tau^+  \tau^-$
within the SM have been estimated to be $B(B\to X_s \tau^+
\tau^-) = 3.2 \times 10^{-7}$ \cite{Hewett:1995dk}.

Using the constraint $|V^{d*}_{Rbb}V^d_{Rbs}|\lsim
3.5\times10^{-3}$ from $\Delta M_{B_s}$ with $\cot\theta_R
\tan\theta_W (m_W/m_{Z^\prime}) \approx 1$ (see
Figure~(\ref{parameter})), we find the following upper bounds for
these decays
\begin{eqnarray}
&&B(B\to X_s \tau^+ \tau^-)\leq 4.4\times 10^{-5},\nonumber\\
&&B(B\to X_s \nu \bar \nu)\leq 3.7\times 10^{-4},\nonumber\\
 &&B(B_s \to \tau \bar
\tau)\leq 6.3\times 10^{-5}. \label{modes}
\end{eqnarray}
These upper bounds are  larger than the respective SM predictions
by factors of about 100, 8 and 55, representing one and two orders
of magnitude enhancements. They occur when the imaginary part of
$V^{d*}_{Rbb}V^d_{Rbs}$ is large (see Figure~(\ref{parameter})).
If we restrict $V^{d*}_{Rbb}V^d_{Rbs}$  to be real, then the
constraint reads $|V^{d*}_{Rbb}V^d_{Rbs}|\lsim 9 \times10^{-4}$
and the largest enhancements possible for the modes in
Eq.~(\ref{modes}) become 10, 1.5 and 5 respectively. The one-loop
new physics contributions are still smaller than this.

Let us now comment on the effect of $\Delta M_{B_s}$ on $K\to \pi
\nu \bar \nu$. Here we would like to see whether the new
constraint on the flavor changing parameters allows for the
enhancement of about 2 over the SM that is required to reproduce
the measured central value $B(K^+ \to \pi^+ \nu \bar \nu) =
(1.47^{+1.30}_{-0.89})\times 10^{-10}$ by E787 and E949
\cite{Adler:2001xv,Artamonov:2004hr}. The tree-level $Z^\prime$
contribution compared with the SM is given by
\begin{eqnarray}
&&{\Gamma_{tree}(K^+\to \pi^+ \nu\bar \nu)\over \Gamma_{SM}(K^+\to
\pi^+ \nu\bar \nu)} \approx  1130 \cot^4\theta_R \tan^4\theta_W
\left({M_W \over M_{Z^\prime}}\right)^4
\left|{V^{d\star}_{Rbs}V^{d}_{Rbd}\over
V^\star_{ts}V_{td}}\right|^2.
\end{eqnarray}

The parameters involved are different than those in $B_s$ mixing.
They can be related when the matrix $V^d_{Rij}$ is almost
diagonal. In that case $V^d_{Rbb} \approx 1$, and
$|V^{d*}_{Rbb}V^d_{Rbs} |\approx |V^d_{Rbs}|\lsim 3.5 \times
10^{-3}$.  A similar analysis for $\Delta M_{B_D}$ using
Figure~(\ref{rangebd}) leads to $|V^d_{Rbd}| \lsim 2.5\times
10^{-4}$. Combining these two results one obtains
\begin{eqnarray}
\left|{V^{d\star}_{Rbs}V^{d}_{Rbd}\over
V^\star_{ts}V_{td}}\right|^2 \lsim 7 \times 10^{-6}
\end{eqnarray}
With these numbers, the tree-level $Z^\prime$ exchange
contributions to $B(K^+ \to \pi^+ \nu \bar \nu)$, $B(B_d \to X_d
\tau^+  \tau^- (\nu\bar \nu))$ and to $B(B_d \to \tau^+ \tau^-
(\nu \bar \nu))$ are much smaller than their SM counterparts.

The situation is different for the one loop level $Z^\prime$
flavor changing interaction Eq.~(\ref{effl}). Here we have
\begin{eqnarray}
{\Gamma_{loop}(K^+\to \pi^+ \nu \bar \nu)\over \Gamma_{SM}(K^+\to
\pi^+\nu\bar \nu)} \approx {1\over 12} \cot^4\theta_R
\tan^4\theta_W \left ({m_W\over m_{Z^\prime}}\right )^4 \left
\vert {I(\lambda_t, \lambda_H)\over X(x_t)}\right \vert^2.
\end{eqnarray}
The total contribution to the rate is simply the sum of the SM and
the one loop $Z^\prime$ exchange since they have the same CKM
factor and the same sign. With $I(\lambda_t, \lambda_H) = 5.54$,
we obtain the central value of $1.47\times 10^{-10}$ by E787 and
E949, and as we saw in Eq.~(\ref{intran}), this value is allowed
by $\Delta M_{B_s}$. A similar situation occurs for the CP
violating decay $K_L \to \pi^0 \nu\bar \nu$ where
\begin{eqnarray}
\Gamma(K_L \to \pi^0 \nu\bar \nu) = \Gamma(K_L \to \pi^0 \nu\bar
\nu)_{SM} {\Gamma(K^+\to \pi^+\nu\bar \nu)\over \Gamma(K^+\to
\pi^+ \nu\bar \nu)_{SM}}.
\end{eqnarray}

\section{CP Violation}

A complex new physics contribution to $M^{B_s,N}_{12}$ can have
significant effects on CP violation in $B_s$ decays \cite{Group1}.
We briefly comment on the effects on two experimental observables,
the dilepton and the time dependent CP asymmetries $a$ and
$A_{TCP}$ defined as
\begin{eqnarray}
&&a = {N^{++} - N^{--}\over N^{++}+N^{--}} =
\frac{\left|\frac{p_{B_s}}{q_{B_s}}\right|^2|A|^4 -
\left|\frac{q_{B_s}}{p_{B_s}}\right|^2 |\bar A|^4 }{
\left|\frac{p_{B_s}}{q_{B_s}}\right|^2|A|^4
+\left|\frac{q_{B_s}}{p_{B_s}}\right|^2 |\bar
A|^4},\nonumber\\
&&A_{TCP} = 2 e^{\frac{\Delta \Gamma^{B_s}}{2} t} {A_f \cos(\Delta
M^{B_s} t) + S_f \sin(\Delta M^{B_s} t)\over 1+ e^{\Delta
\Gamma^{B_s} t} - A^{\Delta \Gamma}_f (1- e^{\Delta \Gamma^{B_s} t})},
\end{eqnarray}
where $N^{ii}$ is proportional to $\Gamma(b \bar b \to l^i l^i X)$
and $A$ and $\bar A$ are the decay amplitudes for $B \to l^+ \nu
X$ and $\bar B \to l^- \bar \nu \bar X$. $\Delta \Gamma$ is the
lifetime difference between the heavy and light states $B_s^L$ and
$B_s^H$. The other quantities are defined as
\begin{eqnarray}
&&A_f = {|A(f)|^2-|\bar A(\bar f)|^2\over |A(f)|^2+|\bar A(\bar
f)|^2},\;\;S_f = - 2 {Im((q_{B_s}/p_{B_s})\bar A(f)A^*(f))\over
|A(f)|^2+|\bar A(\bar f)|^2},\nonumber\\
&&A^{\Delta \Gamma}_f =  2 {Re((q_{B_s}/p_{B_s})\bar
A(f)A^*(f))\over |A(f)|^2+|\bar A(\bar f)|^2},
\;\;|A_f|^2+|S_f|^2+|A^{\Delta \Gamma}_f|^2 = 1. \label{qm}
\end{eqnarray}
Here $A(f)$ and $\bar A(\bar f)$ are decay amplitudes for $B_s$
and $\bar B_s$ decay into CP eigen-states $f$. In terms of the
$B_s$ mixing parameters
\begin{eqnarray}
{q_{B_s}\over p_{B_s}} = \sqrt{{M^{B_s*}_{12} -
i\Gamma^{B_s*}_{12}/2\over M^{B_s}_{12} - i \Gamma^{B_s}_{12}/2}},
\end{eqnarray}

Assuming that CP violation in $A$ and $\bar A$ is small, $|A| =
|\bar A|$, and
\begin{eqnarray}
a = {Im(\Gamma^{B_s}_{12}/M^{B_s}_{12}) \over 1 +
|\Gamma^{B_s}_{12}/M^{B_s}_{12}|^2/4}.
\end{eqnarray}
In the SM one has \cite{Beneke:1998sy}
\begin{eqnarray}
{2\Gamma^{B_s,SM}_{12}\over \Gamma^{B_s}} = -{f^2_{B_s}\over
(230\mbox{MeV})^2}(0.007 B_{B_s} + 0.132 \epsilon - 0.078)\approx
-0.11.
\end{eqnarray}
This number is consistent with the 95\% CL HFAG experimental upper
bound of $\Delta \Gamma(B_s)/\Gamma(B_s) < 0.54$ \cite{hfag}.

Since $\Gamma^{B_s}_{12}= |\Gamma_{12}^{B_s}|e^{i\alpha_s}$ arises from
loop contributions involving light quarks, we do not expect a
significant new physics contribution. In the following we will use
the SM value above to estimate this quantity. In terms of the
phase of $M^{B_s}_{12} = |M_{12}^{B_s}|e^{i\alpha_s + i\theta_s}$,
we obtain
\begin{eqnarray}
a \approx 0.004\sin\theta_s.
\end{eqnarray}
The phase $\theta_s$ is not constrained by the measurement of
$\Delta M_{B_s}$, as we saw it can take any value  from $0$ to
$2\pi$. Therefore the asymmetry $a$ can vary in the range from
-0.004 to 0.004.\footnote{It can also reach 2\% if
$\Delta\Gamma(B_s)/\Gamma(B_s)$ is close to its experimental upper
bound.}

There may also be large effects in the time dependent CP
asymmetry. The time dependent CP asymmetries in B decays have been
shown to provide crucial information about CP violation in $B_d$
decays. For $B_s$ decays, the ``Gold Plated" mode to study CP
violation is $B_s \to \psi \phi$. In the SM, $S_f$ is about 0.038
and $A_f$ is also very small. But with new physics, $S_f$ can be
much larger (since $S_f = \sin(2\theta_s)$) even if there is no CP
violating phase in $A(f)$ and $\bar A(\bar f)$. Future experiments
should test CP violation in the $B_s$ sector \cite{Ball:2000ba}.

There is another special aspect of time dependent CP violation in
$B_s$ decays due to the fact that $\Delta \Gamma$ is not equal to
zero \cite{Ball:2000ba}. If $\Delta \Gamma=0$, which is a very good
approximation for $B_d$ decays, it is not possible to measure
$A^{\Delta \Gamma}_f$, and one cannot check the last equation in
Eq.~(\ref{qm}). Assuming again that CP violation in the decay
amplitudes is small, we have
\begin{eqnarray}
A_{TCP} = 2 e^{\Delta \Gamma^{B_s} t/2} {\sin\theta_s \sin(\Delta
M^{B_s} t)\over 1+ e^{\Delta \Gamma^{B_s} t} - \cos\theta_s (1-
e^{\Delta \Gamma^{B_s} t})}.
\end{eqnarray}

\begin{figure}[htb]
\begin{center}
\includegraphics[width=8cm]{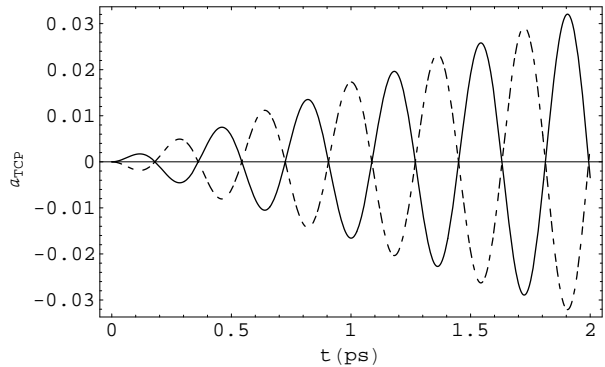}
\end{center}
\caption{$a_{TCP}$ as a function of t (ps) for two values of
$\theta_s$; $\theta_s = 2\pi/3$ shown as a solid line, and $\theta_s
= \pi/5$ shown as a dashed line . }\label{quantf}
\end{figure}

In Figure~(\ref{quantf}), we show $ a_{TCP}=A_{TCP}(\Delta
\Gamma^{B_s}) - A_{TCP}(0)$ as a function of t. We have chosen two
values $2\pi/3$ and $\pi/5$ for $\theta_s$ for illustration. We
can see that at a few percent level, there are differences
compared with the $\Delta \Gamma =0$ case such difference may be
tested at LHCb.

\noindent {\bf Acknowledgments}$\,$ The work of X.G.H. was
supported in part by the National Science Council under NSC
grants. The work of G.V. was supported in part by DOE under
contract number DE-FG02-01ER41155. We thank Soeren Prell for
useful conversations.

\end{document}